\documentclass[12pt]{article}

\usepackage{enumitem}
\usepackage{graphicx}
\usepackage{float}
\usepackage{cite}
\usepackage{amsfonts}
\usepackage{amssymb}
\usepackage{amsmath}
\usepackage{xcolor}

\def\gtwid{\mathrel{\raise.3ex\hbox{$>$\kern-.75em\lower1ex\hbox{$\sim$}}}}
\def\ltwid{\mathrel{\raise.3ex\hbox{$<$\kern-.75em\lower1ex\hbox{$\sim$}}}}
\def\square{\kern1pt\vbox{\hrule height 1.2pt\hbox{\vrule width 1.2pt\hskip 3pt
  \vbox{\vskip 6pt}\hskip 3pt\vrule width 0.6pt}\hrule height 0.6pt}\kern1pt}

\begin{document}

\begin{titlepage}

\begin{flushright}
CCTP-2025-09 \\
ITCP/2025/09 \\
UFIFT-QG-25-04
\end{flushright}

\vskip 1cm

\begin{center}
{\bf Quantum Cosmology in Accelerating Spacetimes}
\end{center}

\vskip 1cm

\begin{center}
S. P. Miao$^{1\star}$, N. C. Tsamis$^{2\dagger}$ and 
R. P. Woodard$^{3\ddagger}$
\end{center}

\vskip 0.5cm

\begin{center}
\it{$^{1}$ Department of Physics, National Cheng Kung University, \\
No. 1 University Road, Tainan City 70101, TAIWAN}
\end{center}

\begin{center}
\it{$^{2}$ Institute of Theoretical Physics \& Computational Physics, \\
Department of Physics, University of Crete, \\
GR-700 13 Heraklion, HELLAS}
\end{center}

\begin{center}
\it{$^{3}$ Department of Physics, University of Florida,\\
Gainesville, FL 32611, UNITED STATES}
\end{center}

\vspace{0cm}

\begin{center}
ABSTRACT
\end{center}
We simplify the gravitational equations which apply 
in accelerating spacetimes and are consistent with 
the cosmological principle. Solutions to these equations 
should be tantamount to all order re-summations of the 
perturbative leading logarithms. We discuss the ``null
hypothesis'' and we study the local expansion rate
observable.

\begin{flushleft}
\end{flushleft}

\vskip 0.5cm

\begin{flushleft}
$^{\star}$ e-mail: spmiao5@mail.ncku.edu.tw \\
$^{\dagger}$ e-mail: tsamis@physics.uoc.gr \\
$^{\ddagger}$ e-mail: woodard@phys.ufl.edu
\end{flushleft}

\end{titlepage}

\section{Prologue}

A prototypical example of an accelerating spacetime becoming 
an important physical paradigm during early cosmological 
evolution is the primordial inflationary era defined by 
$H > 0$ with $0 \leq \epsilon < 1$; any cosmological geometry 
is characterized by a scale factor $a(t)$ and its two first 
time derivatives, the Hubble parameter $H(t)$ and the 1st slow 
roll parameter $\epsilon(t)$:
\footnote{It is more often than not convenient to employ 
conformal instead of co-moving coordinates:
$ds^2 \!=\! 
-dt^2 + a^2(t) \, d{\mathbf{x}} \cdot d{\mathbf{x}} 
= 
a^2(\eta) \big[\! -d\eta^2 + a^2(\eta) \, 
d{\mathbf{x}} \cdot d{\mathbf{x}} \big]$, with
$t$ the co-moving time and $\eta$ the conformal
time.}
\begin{equation}
ds^2 = - dt^2 + a^2(t) \, d{\mathbf{x}} \cdot d{\mathbf{x}} 
\qquad \Longrightarrow \qquad 
H(t) \equiv \frac{\dot{a}}{a} 
\quad, \quad 
\epsilon(t) \equiv - \frac{\dot{H}}{H^2} 
\; . \label{geometry}
\end{equation}

Because gravitation plays the dominant role in shaping the 
cosmological evolution, it is within the gravitational sector
that one should {\it first} look for a natural ``inflation 
causing mechanism''. Indeed the presence of a bare and positive 
cosmological constant $\Lambda$ provides such a mechanism; 
de Sitter spacetime is a solution of the field 
equations.\footnote{Although it shall be rather apparent 
in our analysis what is valid for any accelerating spacetime, 
it will be the de Sitter geometry that is exclusively used 
to obtain our results.}
Furthermore, this is a very natural situation since $\Lambda$ 
is constant in {\it space} and no special initial condition 
is needed to start inflation. The physical implications
of this mechanism in a theory notorious for complicated 
calculations have been investigated to some degree in 
perturbation theory \cite{Tsamis:1994ca} and in Section 2 
we present the elements of quantum gravity necessary for 
our study.

A straightforward application of quantum field theoretic methods 
to the gravitational effective theory of Section 2 shows the 
presence of factors of $\ln[a(t)]$ in correlators which in a 
long period of inflation can grow large enough to overwhelm 
even the smallest coupling constant causing perturbation theory 
to break \cite{Woodard:2025cez}. To overcome this, a stochastic 
technique that re-sums the leading of these logarithmic factors 
(LLOG) was pioneered by Starobinsky \cite{Starobinsky:1986fx} 
and, later, was extended to include theories with derivative 
interactions like gravitation \cite{Miao:2024shs,Miao:2025gzm}. 
The final form of the LLOG gravitational equations that are 
appropriate for accelerating spacetimes can be found in 
Section 3.

In Section 4, the LLOG equations of Section 3 are applied 
to (in)validate the ``null hypothesis'' which states that
besides the tensor CMBR spectrum there are {\it no} 
significant gravitational effects during the inflationary era.
 
An appropriate observable to evaluate given our LLOG equations 
is the one corresponding to the physical expansion rate.
This is constructed and thereafter analyzed in Section 5.

Finally, in the Epilogue we discuss the physical implications 
and prospects from this study, and in the Appendix we display 
the matrices of the Arnowitt, Deser, Misner (ADM) spatial-temporal 
decomposition.

\section{Quantum Gravity}

The most generic gravitational effective theory consistent
with general coordinate invariance can be organized as a
infinite series of terms with increasing canonical dimensionality.
For cosmological considerations the two lowest terms should
suffice:
\footnote{Hellenic indices take on spacetime values
while Latin indices take on space values. Our metric
tensor $g_{\mu\nu}$ has spacelike signature
$( - \, + \, + \, +)$ and our curvature tensor equals
$R^{\alpha}_{~ \beta \mu \nu} \equiv
\Gamma^{\alpha}_{~ \nu \beta , \mu} +
\Gamma^{\alpha}_{~ \mu \rho} \,
\Gamma^{\rho}_{~ \nu \beta} -
(\mu \leftrightarrow \nu)$.}
\begin{equation}
{\mathcal L}_{inv} =
\frac{1}{\kappa^2} \Big[ \!-\! (D \!-\! 2) \Lambda + R \, \Big] \sqrt{-g}
\; , \label{Linv}
\end{equation}
where the two parameters are Newton's constant $G$ and the 
cosmological constant $\Lambda$:
\footnote{$\Lambda$ is taken to be ``large'' and positive. 
Here ``large'' means a $\Lambda$ corresponding to a scale
$M$ which can be as high as $10^{18} \, GeV$.} 
\begin{equation}
\kappa^2 \equiv 16 \pi G
\quad , \quad
\Lambda \equiv (D \!-\! 1) H^2
\; . \label{parameters}
\end{equation}

The full equations coming from (\ref{Linv}) are:
\begin{equation}
R_{\mu\nu} - \frac12 R \, g_{\mu\nu} 
+ \frac12 (D \!-\! 2)(D \!-\! 1) H^2 g_{\mu\nu} = 0
\; . \label{eom}
\end{equation}

One could argue that the proper methodology to study 
(\ref{Linv}) is to consider it as a {\it quantum} effective
effective field theory (QEFT) to which the full complement
of quantum field theory (QFT) can be applied.\footnote{This 
has been emphasized in many and various occasions by 
Weinberg in his approach to the most general effective 
field theory, for instance in \cite{DHoker:1994rdl}.}
The argument that here we are dealing with a 
non-renormalizable theory is of no real physical 
significance. The Bogoliubov-Parasiuk-Hepp-Zimmermann
(BPHZ) procedure will always eliminate the perturbative 
divergences in loop amplitudes; the {\it only} difference 
is that for non-renormalizable theories there is loss of 
{\it predictability} due to the ever-increasing number of 
counterterms with undetermined coupling constants. However, 
in our case it is only the terms present in (\ref{Linv}) 
that are relevant and their coupling constants are physically 
determined.

In terms of the full metric $g_{\mu\nu}$, the conformally
re-scaled metric ${\widetilde g}_{\mu\nu}$ and the graviton 
field $h_{\mu\nu}$ are defined thusly:
\begin{equation}
g_{\mu\nu} \equiv 
a^2 {\widetilde g}_{\mu\nu} \equiv
a^2 \big[ \eta_{\mu\nu} + \kappa h_{\mu\nu} \big] 
\; . \label{metrics}
\end{equation}

The requirement of complete calculability needs appropriate:
\\ [3pt]
{\it (i) regularization:}
We use dimensional regularization, a well-established 
technique that respects general co-ordinate invariance 
and includes {\it all} available modes. One merely 
{\it assumes} that as the high frequency modes redshift 
they equilibrate into the gravitational field equations 
of general relativity.
\\ [3pt]
{\it (ii) gauge fixing condition:} It would be clearly 
desirable to employ a de Sitter invariant gauge condition 
but unfortunately the resulting propagator would be very 
complicated. However, one could still impose an exact 
de Sitter invariant gauge in which case the propagator 
equation would be de Sitter invariant but would not 
possess a de Sitter invariant solution.\footnote{The 
classic example is the theory of a massless minimally 
coupled scalar in de Sitter background 
\cite{Allen:1987tz,Miao:2010vs}. Another more recent 
example comes from the computation of the graviton 
1-loop correction to the vacuum polarization 
\cite{Glavan:2015ura}.} Moreover, the de Sitter invariant 
solutions of the graviton propagator equation are 
problematical because they do not correspond to proper
propagators; not {\it every} Green's function is a 
propagator.\footnote{For instance, the ``propagator'' 
proposed in \cite{Morrison:2013rqa} although de Sitter 
invariant, not only lacks calculability but also does 
not correspond, as it should, to the time-ordered product 
of two graviton fields \cite{Miao:2013isa}.} Provided 
that physical observables are computed, it is perfectly 
adequate to employ the graviton propagator that allows 
regularized, BPHZ-renormalized loop calculations to be 
fully completed, even if it breaks de Sitter invariance. 
\footnote{This is accomplished by virtue of its 
{\it coordinate independent} tensor structure and 
the presence of only two ``scalar propagator'' terms 
in $D \!=\! 4$.}

\newpage

\section{The leading logarithm all orders equations}

If the purpose is to evaluate gravitational observables
in an accelerating spacetime like the de Sitter geometry
and if this is to be done within the notoriously complicated
theory of quantum gravity, we must have accumulated strong 
evidence that there are strong gravitational effects 
potentially present. For if this is the case, the universally 
attractive nature of the gravitational interaction {\it may}
alter cosmological parameters, kinematical parameters and 
long-range forces; and such a possibility makes the whole
effort worthwhile

The dimensionless coupling constant of (\ref{Linv}) 
$G \Lambda \!\sim\! G H^2$ stays very small even for 
$M \!=\! 10^{18} \, GeV$ because even then $G \Lambda 
\!=\! (\tfrac{M}{M_{\rm Pl}})^4 \!\sim\! 10^{-4}$ 
and hence is a sensible perturbative parameter.
In the de Sitter background, as time evolves more 
and more real quanta pairs emanate from the vacuum 
\cite{Schrodinger:1939}, and correlators involving 
interacting quanta lead to the theory explicitly 
displaying a secular growth by powers of $\ln[a(t)]$ 
which eventually overwhelms $G H^2$ causing perturbation 
theory to break \cite{Woodard:2025cez}.\footnote{Explicit 
examples of this particular secular growth behaviour can 
be found in citations [14,16,24,31,32,38,44,72,73,77,78] 
contained in reference \cite{Woodard:2025cez}.}
Terms with the maximum power of $\ln(a)$ at a
given perturbative order are known as {\it leading logarithm}
(LLOG) contributions; those with fewer factors of $\ln(a)$
are known as {\it subleading logarithm} contributions.
In general some of these logarithms derive from the 
renormalization procedure while others come from the ``tail'' 
logarithmic term in the graviton propagator. The latter can 
be described by a variant of Starobinsky's stochastic 
formalism \cite{Miao:2025gzm} which is the natural method 
to re-sum them.

The geometrical variables are those of the ADM spatial-temporal 
decomposition of the conformally re-scaled metric 
\cite{Arnowitt:1962hi}:
\begin{eqnarray}
d{\widetilde s}^{\,\,\! 2} 
& \!\!\!=\!\!\! & 
- N^2 d\eta^2 + \gamma_{ij} (dx^i - N^i d\eta) (dx^j - N^j d\eta)
\; , \label{3+1a} \\
& \!\!\!=\!\!\! & 
(- N^2 \!+\! \gamma_{ij} N^i N^j) \, d\eta^2
- 2 \gamma_{ij} N^j d\eta \, dx^i + \gamma_{ij} dx^i dx^j
\; , \label{3+1b}
\end{eqnarray}
where $N$ is the lapse function, $N^i$ is the shift function, 
and $\gamma_{ij}$ is the spatial metric. Furthermore, from 
the spatial-temporal matrices (\ref{3+1lower}-\ref{3+1upperB}) 
in the Appendix, ${\overline \gamma}^{\mu\nu}$ is seen to be 
the ``spatial part'' and $u_{\mu}$ the ``temporal part''.

The graviton field variables are given by the following 
convenient form:
\begin{equation}
\kappa \, h_{\alpha\beta}
\equiv 
A_{\alpha\beta}
+ u_{(\alpha} B_{\beta)}
+ ( u_{\alpha} u_{\beta} \!+\! \overline{\gamma}_{\alpha\beta} ) C
\quad , \quad
u^{\alpha} A_{\alpha\beta} = u^{\alpha} B_{\alpha} = 0
\; , \label{ABC}
\end{equation}
where $A_{\alpha\beta}$ contains the dynamical degrees 
of freedom, $B_{\alpha}$ is canonically associated with 
the momentum constraints of general relativity and physically 
represents the strictly general relativistic potentials, 
and $C$ is canonically associated with the Hamiltonian 
constraint of general relativity and physically represents 
the Newtonian potential. 
\footnote{In addition, $C$ is the constrained field variable 
in which any dynamical changes of the geometrical background 
are imprinted.}

The field variables $A_{ij}, B_i, C$ are related to 
the geometrical variables $\gamma_{ij}, N^i, N$ thusly
\cite{Miao:2025gzm}:
\begin{eqnarray}
N^2 
& \!\!\!=\!\!\! &
\frac{1}{1 + C + \tfrac14 B_i B_i}
\; , \label{N} \\
N^i 
& \!\!\!=\!\!\! &
\frac{\tfrac12 B_i}{\sqrt{1 + C + \tfrac14 B_i B_i}}
\; , \label{Ni} \\
\gamma_{ij}
& \!\!\!=\!\!\! &
\frac{\delta_{ij} + A_{ij}}{1 - C}
\; . \label{gamma_ij} 
\end{eqnarray}

The stochastic rules appropriate to pure gravity can be 
found in \cite{Miao:2024shs} and a more compact review 
of them in \cite{Miao:2025gzm}. They can be summarized 
in two expressions: {\it ``stochastic reduction''} 
(the original rule for non-derivative interactions) 
and {\it ``integrating out''} (the necessary additional 
rule for derivative interactions). The stochastic rule 
has been established for quite some time, however its 
application range is restricted to theories without 
derivative interactions. The integrating out extra rule 
for theories with derivative interactions is the result
of a very long process consisting of sequential studies 
of general scalar potential models \cite{Tsamis:2005hd}, 
Yukawa theory \cite{Miao:2006pn}, scalar electrodynamics 
\cite{Prokopec:2007ak}, non-linear sigma models 
\cite{Miao:2021gic}, and massless minimally coupled scalar 
corrections to gravity \cite{Miao:2024nsz}; all in
de Sitter background. These studies included perturbative 
calculations at 1-loop and 2-loop orders which provide 
very stringent correspondence limits for any proposed 
LLOG re-summation rule. In all the above we anchor our 
formalism in quantum field theory and, except for the 
current work on quantum gravity, confirm their correctness 
against explicit perturbative results.

The rationale for integrating out differentiated 
field bilinears in the presence of a constant graviton 
background is as follows. Differentiated free fields 
cannot produce logarithms, however, they do give rise 
to interactions between undifferentiated fields. 
Integrating the differentiated graviton bilinears 
in the presence of constant graviton background 
captures these interactions to {\it all} orders 
in perturbation theory.

In \cite{Miao:2025gzm} we derived the following form 
of the LLOG equations for pure gravity: 
\begin{eqnarray}
{\dot A}^{\mu\nu} \!-\! \mathcal{\dot A}^{\mu\nu} 
& \!\!\!=\!\!\! & 
\tfrac43 \frac{\kappa^2 \widetilde{H}^3}{8 \pi^2} 
\, \overline{\gamma}^{\mu\nu} 
\Big[ 2 \!-\! \overline{\gamma}^{\alpha\beta} A_{\alpha\beta} 
\!-\! 4C \Big] 
- \tfrac23 \, a^{-1} \, \overline{\gamma}^{\rho (\mu} A^{\nu) \sigma} 
\partial_{\rho} B_{\sigma} 
\; , \qquad \label{Afinal2} \\
B^{\mu} 
& \!\!\!=\!\!\! &
\frac{\kappa^2 \widetilde{H}^2}{16 \pi^2} B^{\mu}  
- 2 A^{\mu\nu} B_{\nu} 
\; , \label{Bfinal2} \\
C 
& \!\!\!=\!\!\! & 
\frac{\kappa^2 \widetilde{H}^2}{8 \pi^2} 
\Big[ - 1 \!+\! 2 \overline{\gamma}^{\alpha\beta} A_{\alpha\beta} 
\!+\! \tfrac72 C \Bigr] 
\; . \label{Cfinal2}
\end{eqnarray}

We can further simplify equations (\ref{Afinal2}-\ref{Cfinal2})
by taking into account the physical arena in which they are 
to operate. For example, if we seek to describe the homogeneous 
expansion history of the universe, it seems apparent that we 
can set $B_i \!=\! 0$ by appealing to the isotropy requirement 
of the cosmological principle. Moreover, assuming our metric
tensor must not diverge, the constrained variable $C$ is 
restricted to be of $O(1)$ as can be readily seen from equation 
(\ref{N}) -- where as $C \!\rightarrow\! -1$ the lapse $N$ 
diverges -- and from equation ({\ref{gamma_ij}) -- where as 
$C \!\rightarrow\! +1$ the metric $\gamma_{ij}$ diverges.
\footnote{The last term in (\ref{Cfinal2}) is sub-dominant
because $\; C \, \big( 1 \!-\! 
\tfrac72 \tfrac{\kappa^2 \widetilde{H}^2}{8 \pi^2} \big)
= 
\tfrac{\kappa^2 \widetilde{H}^2}{8 \pi^2} \big[ 
\!-\! 1 \!+\! 2 \overline{\gamma}^{\alpha\beta} A_{\alpha\beta}
\big]$ and 
$\; \tfrac{\kappa^2 \widetilde{H}^2}{8 \pi^2} \leq 10^{-10}$.
The same reasoning explains why the $-4C$ term in (\ref{Afinal2}) 
is ignored.}

The result of this final reduction is a remarkably simple final 
form for the LLOG gravitational equations in accelerating constant 
background spacetimes:
\begin{eqnarray}
{\dot A}_{ij} \!-\! \mathcal{\dot A}_{ij} 
& \!\!\!=\!\!\! & 
\tfrac43 \frac{\kappa^2 \widetilde{H}^3}{8 \pi^2} 
\, \gamma_{ij} 
\Big[ 2 \!-\! \gamma^{rs} A_{rs} \Big] 
\; , \label{Afinal3} \\
B_i 
& \!\!\!=\!\!\! &
0
\; , \label{Bfinal3} \\
C 
& \!\!\!=\!\!\! & 
\frac{\kappa^2 \widetilde{H}^2}{8 \pi^2} 
\Big[ \!-\! 1 \!+\! 2 \gamma^{ij} A_{ij} \Bigr] 
\; . \label{Cfinal3}
\end{eqnarray}
The above set of equations (\ref{Afinal3}-\ref{Cfinal3})
-- one differential and two algebraic -- contain the 
leading logarithms from all orders of perturbation theory 
and can be seen to be a {\it huge} simplification of the 
original field equations of pure gravity coming from 
$\mathcal{L}_{tot} = \mathcal{L}_{inv} 
+ \mathcal{\widetilde L}_{GF} 
+ \mathcal{\widetilde L}_{gh}$ \cite{Miao:2025gzm}.
It is worth noting that the +2 constant term in 
(\ref{Afinal3}), as well as the -1 constant term 
in (\ref{Cfinal3}) furnish very small contributions 
$( \tfrac{\kappa^2 \widetilde{H}^2}{8 \pi^2} 
\!\leq\! 10^{-10} )$ which are clearly sub-dominant to 
the term involving the growing dynamical field $A_{ij}$.

To extract physical information from the operator equations 
(\ref{Afinal3}-\ref{Cfinal3}), the straightforward way would 
be to take the vacuum expectation value (VEV) of a physical 
observable and obtain its complete (non-perturbative) LLOG 
time evolution; this is still a non-trivial project to say 
the least but now it is a project which has perhaps become 
a rather realistic endeavour. In terms of the physical 
picture coming out of the system of equations 
(\ref{Afinal3}-\ref{Cfinal3}), the constrained fields start 
very small (or zero) and grow very slowly (or stay zero) 
as they {\it respond} to the growth of the dynamical field. 

As an additional bonus, there is another way to use the
operator equations (\ref{Afinal3}-\ref{Cfinal3}). The
perturbative order by order limit of the complete LLOG
result permits to easily determine the leading logarithm
result from loop calculations which would normally take a 
long long time to conclude.\footnote{Starobinsky exploited 
both ways to obtain both non-perturbative and perturbative 
results in the late time limit for his simple self-interacting 
scalar field model \cite{Starobinsky:1986fx,Tsamis:2005hd}.}
Furthermore, since by their mathematical structure the 
LLOG equations should re-sum all the perturbative leading 
logarithms, they should have a range of validity far 
beyond that of perturbation theory, all the way until 
the asymptotic solution is reached.

\section{The null hypothesis}

- A hypothesis whose validity our LLOG equations should 
be able to discern is the ``null hypothesis'' which states 
that during the inflationary epoch no quantum gravitational 
effects were present other than the tensor primordial power 
spectrum. Indeed, it is the long-wavelength graviton production 
during inflation which is thought to be the physical mechanism
causing this spectrum and this can be realized by the presence 
of a stochastic free field jitter $\mathcal{A}_{ij}$. 

The background free field $\mathcal{A}_{ij}$ commutes with 
itself and has a free field expansion in terms of creation 
and destruction operators which can take any value making 
$\mathcal{A}_{ij}$ a random variable, in other words a 
stochastic field:
\begin{equation}
\mathcal{A}_{ij}(t, {\bf x}) =\!
\int_{k=H}^{H a(t)} \!\! \frac{d^3k}{(2\pi)^3}
\frac{H}{\sqrt{2k^3}}
\Big[ e^{i {\bf k} \cdot {\bf x}} \, \alpha_{ij}({\bf k})
+ e^{-i {\bf k} \cdot {\bf x}} \, \alpha^{\dagger}_{ij}({\bf k}) \Big]
\; . \label{stochfield}
\end{equation}
The reason (\ref{stochfield}) exhibits classical 
behaviour by commuting with itself is because both 
the creation and destruction part are multiplied by 
the identical mode function $\tfrac{H}{\sqrt{2k^3}}$, 
i.e. by the long-wavelength (IR) limit of the de 
Sitter background; nonetheless, $\mathcal{A}_{ij}$ 
still is an operator field.

In terms of the LLOG equations (\ref{Afinal3}-\ref{Cfinal3}),
the null hypothesis implies the following solution:
\begin{equation}
``null \; hypothesis"
\quad \Longrightarrow \quad
A_{ij} = \mathcal{A}_{ij}
\;\; \& \;\;
B_i = C = 0
\; , \label{null}
\end{equation}
which is the pure de Sitter geometry.

In view of (\ref{null}), equation (\ref{Afinal3})
becomes:
\begin{equation}
0 = \tfrac43 \frac{\kappa^2 \widetilde{H}^3}{8 \pi^2}
\, \gamma_{ij}
\Big[ 2 \!-\! \gamma^{rs} A_{rs} \Big]
\; , \label{nullAeqn}
\end{equation}
The factor of 2 inside the square brackets of (\ref{nullAeqn}) 
would be enough to falsify the null hypothesis, except that it 
could be eliminated by a finite renormalization of $\Lambda$. 
However, the second term {\it perturbatively} goes like 
$-\kappa^2 H^2 \ln(a)$ for at least as long as 
$\kappa^2 H^2 \ln(a) \!\ll\! 1$, and this {\it suffices 
to falsify} the null hypothesis.

That said, by algebraically manipulating $\gamma^{rs} A_{rs}$:
\begin{eqnarray}
\gamma^{rs} A_{rs} 
& \!\!\!=\!\!\! &
\gamma^{rs} \mathcal{A}_{rs}
=
\big( \tfrac{1}{I + \mathcal{A}} \big)^{rs} \mathcal{A}_{sr}
=
{\rm Tr} \Big[ \tfrac{1}{I + \mathcal{A}} \, \mathcal{A} \Big]
\; , \nonumber \\
& \!\!\!=\!\!\! &
{\rm Tr} \Big[ \big( I - \mathcal{A} + \mathcal{A}^2 - \mathcal{A}^3
+ \mathcal{A}^4 - \dots \big) \mathcal{A} \Big]
\; , \nonumber \\
& \!\!\!=\!\!\! &
+ \, {\rm Tr} \Big[ \mathcal{A} + \mathcal{A}^3 + \mathcal{A}^5 
+ \dots \Big]
- {\rm Tr} \Big[ \mathcal{A}^2 + \mathcal{A}^4 + \mathcal{A}^6 
+ \dots \Big]
\; , \label{gamma A}
\end{eqnarray}
we can separate it into an odd and an even series of traces.
The VEV of the odd series vanishes because $\mathcal{A}_{ij}$
is a free field and its odd powers contain odd numbers of
creation and destruction operators. The VEV of the even series
gives a series of contributions increasing with time.

Because the null hypothesis is incorrect there are physical 
gravitational effects present during the inflationary period 
beyond the tensor primordial power spectrum. Moreover the 
LLOG equations (\ref{Afinal3}-\ref{Cfinal3}) evidently do 
not allow for a free solution nor do they allow for eternal 
de Sitter geometry.

\newpage

\section{The expansion rate observable}

Because (\ref{Afinal3}-\ref{Cfinal3}) are {\it operator} 
equations, their gauge dependence can be eliminated by 
inserting the solutions in a {\it gauge invariant} operator.
A physical measure of gravitational back-reaction can be 
provided by an observable which invariantly determines 
the expansion rate \cite{HawkEll}.

The standard local definition of the expansion rate
${\cal H}$ in $D$ dimensions:
\begin{equation}
{\cal H}(t, {\bf x})
\, = \,
- \tfrac{1}{D-1} \, {\cal D}^{\mu} V_{\mu}(t, {\bf x})
\;\; , \label{H}
\end{equation}
is in terms of the covariant derivative ${\cal D}^{\mu}$ 
of a timelike $D$-velocity field $V_{\mu}$:
\begin{equation}
g^{\mu\nu}(x) \, V_{\mu}(x) \, V_{\nu}(x)
\, = \, - 1
\; . \label{V}
\end{equation}
An appropriate $D$-velocity field can be constructed from a 
scalar functional of the metric $\varphi[g](x)$ satisfying, 
for all $x$, the following {\it first} order non-linear
dynamical equation:
\footnote{This choice was motivated from 
\cite{Geshnizjani:2002wp, Deffayet:2024ciu} where the use 
of equation (\ref{varphi}) allowed simpler calculations
compared to choosing a {\it second} order equation 
\cite{Tsamis:2014kda}. Further approaches to invariant 
expansion observables can also be found in 
\cite{Finelli:2011cw, Marozzi:2012tp, Marozzi:2014xma}.}
\begin{equation}
g^{\mu\nu} \partial_{\mu} \varphi \, \partial_{\nu} \varphi 
\, = \, - 1
\; . \label{varphi}
\end{equation}
On the initial value surface the scalar functional satisfies:
\begin{equation}
\varphi(t=0, {\bf x}) \Big\vert_{\rm IVS} \, = \, 0
\; . \label{IVD}
\end{equation}
The resulting expansion rate according to (\ref{H}) is: 
\begin{eqnarray}
{\cal H}[g](x) 
& \!\!\!=\!\!\! &
- \tfrac{1}{D-1} \, {\cal D}^{\mu} V_{\mu}[g](x)
=
- \tfrac{1}{D-1} \, {\cal D}^{\mu} \partial_{\mu} \varphi[g](x)
\; , \label{H2a} \\ 
& \!\!\!=\!\!\! & 
- \tfrac{1}{(D-1) \, \sqrt{-g}} \, \partial_{\mu} \big[ 
\sqrt{-g} \, g^{\mu\nu} \, \partial_{\nu} \varphi \big]
\; . \label{H2b}
\end{eqnarray}
Under general coordinate transformations which preserve 
the initial value surface, the variable just constructed 
transforms thusly:
\begin{equation}
{\cal H}[g'] (x) \, = \, {\cal H}[g] (x'^{\, -1}(x))
\; . \label{transf}
\end{equation}

We can invariantly fix the observation time by specifying
the surfaces of simultaneity as follows:
\begin{equation}
\varphi[g](\vartheta [g](x), {\bf x}) \, = \,
\varphi_{\rm dS} (t)
\;\; , \label{time}
\end{equation}
where $\varphi_{\rm dS} (t)$ is the scalar $\varphi$ in de 
Sitter spacetime. This requirement determines the functional
$\vartheta [g](x)$ or, equivalently, the observation time.
Because on de Sitter spacetime he unique solution for the 
scalar is $\varphi(t,{\bf x}) \!=\! t$, we get a {\it full 
invariant} by evaluating: 
\begin{equation}
\mathcal{H}[g](t,{\bf x}) \;\, {\rm at} \;\, 
t \!=\! \varphi[g](t,{\bf x})
\quad , \quad
{\mathcal H}[g'] (t, {\bf x}) =  
{\mathcal H}[g] (t, x'^{\, -1}(t, {\bf x}))
\; . \label{Hinv}
\end{equation}

We now use the ADM representation (\ref{3+1a}-\ref{3+1b}) 
for the spatial-temporal decomposition of the non-linear 
equation (\ref{varphi}):
\footnote{Here and henceforth a prime denotes conformal 
time differentiation $\partial_0$ while a dot stands for 
differentiation with respect to co-moving time $\partial_t$, 
e.g. $\varphi' \! = a \ {\dot \varphi}$.} 
\begin{eqnarray}
-1 = g^{\mu\nu} \partial_{\mu} \varphi \, \partial_{\nu} \varphi
& \!\!\!=\!\!\! &
g^{00} (\varphi')^2 + 2 g^{0i} \varphi' \, \partial_i \varphi 
+ g^{ij} \partial_i \varphi \, \partial_j \varphi 
\; , \label{varphi3+1a} \\
& \!\!\!=\!\!\! & 
- \tfrac{1}{a^2 N^2} \big( \varphi' + N^i \partial_i \varphi \big)^2
+ \tfrac{1}{a^2} \gamma^{ij} \partial_i \varphi \, \partial_j \varphi
\; , \label{varphi3+1b}
\end{eqnarray}
and therefore solve for the co-moving time derivative 
of $\varphi$ in terms of the geometrical variables, and 
by using (\ref{N}-\ref{gamma_ij}) in terms of the field 
variables:
\begin{eqnarray}
\dot{\varphi} 
& \!\!\!=\!\!\! &
N \Big\{ \sqrt{ 1 + \gamma^{ij} 
\big( \tfrac{\partial_i \varphi}{a} \big)
\big( \tfrac{\partial_j \varphi}{a} \big) }
- \tfrac{N^i}{N} \big( \tfrac{\partial_i \varphi}{a} \big) \Big\}
\; , \label{varphidot1} \\
& \!\!\!=\!\!\! &
\tfrac{1}{\sqrt{1 + C + \tfrac14 B_r B_r}} \Big\{
\sqrt{ 1 + (1 - C) \big( \tfrac{1}{I + A} \big)^{\! ij} 
\big( \tfrac{\partial_i \varphi}{a} \big)
\big( \tfrac{\partial_j \varphi}{a} \big) }
- \tfrac{1}{2} B_i \big( \tfrac{\partial_i \varphi}{a} \big) \Big\}
\; . \qquad \label{varphidot2}
\end{eqnarray}

Moreover, we can compute the scalar measure of the 
expansion rate $\cal{H}$ starting from (\ref{H2b})
and using results following from (\ref{3+1b}):
\footnote{For instance, 
$\varphi' \!+\! N^i \partial_i \varphi = 
a N \sqrt{ 1 + \gamma^{ij} 
\big( \tfrac{\partial_i \varphi}{a} \big) 
\big( \tfrac{\partial_j \varphi}{a} \big) }$.}
\begin{eqnarray}
{\cal H}[g](x) 
& \!\!\!=\!\!\! & 
- \tfrac{1}{(D-1) \, a^D \sqrt{-\widetilde{g}}} \, 
\partial_{\mu} \big[ a^{D-2} \sqrt{-\widetilde{g}} \, 
{\widetilde g}^{\mu\nu} \, \partial_{\nu} \varphi \big]
\; , \label{H3a} \\
& \!\!\!=\!\!\! &
- \tfrac{1}{(D-1) \, a^D N \sqrt{\gamma}} \Bigl\{ 
\partial_0 \Bigl[ a^{D-2} N \sqrt{\gamma} \, \big( 
\widetilde{g}^{00} \phi' \!+\! 
\widetilde{g}^{0j} \partial_j \varphi \big) \Bigr] 
\nonumber \\
& &
+ \, \partial_i \Bigl[ a^{D-2} N \sqrt{\gamma} \, \big(
\widetilde{g}^{i0} \varphi' \!+\! 
\widetilde{g}^{ij} \partial_j \varphi \big) \Bigr] \Bigr\} 
\; , \label{H3b} \\
& \!\!\!=\!\!\! &
+ \, \tfrac{1}{(D-1) \, a^D N \sqrt{\gamma}} \Big\{ 
\partial_0 \Big[ a^{D-2} \tfrac{1}{N} \sqrt{\gamma} \big( 
\varphi' \!+\! N^i \, \partial_i \varphi \big) \Big]  
\nonumber \\
& &
+ \, \partial_i \Big[ a^{D-2} \tfrac{1}{N} \sqrt{\gamma} \, N^i \big(
\varphi' \!+\! N^j \, \partial_j \varphi \big) \Big]
- \partial_i \Big[ a^{D-2} N \sqrt{\gamma} \, \gamma^{ij} 
\partial_j \varphi \Big] \Big\}
\; , \qquad \label{H3d} \\
& \!\!\!=\!\!\! & 
\tfrac{H}{N} \sqrt{ 1 + \gamma^{rs} 
\big( \tfrac{\partial_r \varphi}{a} \big)
\big( \tfrac{\partial_s \varphi}{a} \big) }
+ \tfrac{1}{(D-1) N \sqrt{\gamma}} \Big\{
\!\!-\! \tfrac{1}{a} \, \partial_i \big[ N \sqrt{\gamma} \, 
\gamma^{ij} \big( \tfrac{\partial_j \varphi}{a} \big) \big]
\nonumber \\
& & 
+ \bigl( \partial_t \!+\! \tfrac{1}{a} \partial_i \, N^i \bigr) 
\Big[ \sqrt{\gamma} \, 
\sqrt{ 1 \!+\! \gamma^{rs} \big( \tfrac{\partial_r \varphi}{a} \big)
\big( \tfrac{\partial_s \varphi}{a} \big) } \Big] \Big\} 
\; . \label{H3e}
\end{eqnarray}
At late times it makes sense to ignore the spatial derivative 
terms, which redshift like $a^{-1}$. This gives:
\begin{equation}
late \; times: \quad 
\mathcal{H} \longrightarrow \tfrac1{N} \Bigl\{ H + \tfrac{\gamma^{ij} 
\dot{\gamma}_{ij}}{2 (D-1)} \Bigr\} 
\sim
\tfrac1{N} \Bigl\{ H + \tfrac{\gamma^{ij} \dot{A}_{ij}}{2 (D-1)} \Bigr\}
\; . \label{Hlate}
\end{equation}
An algebraic exercise similar to (\ref{gamma A}):
\begin{eqnarray}
\gamma^{ij} {\dot A}_{ij} 
& \!\!\!=\!\!\! &
\big( \tfrac{1}{I + A} \big)^{ij} {\dot A}_{ij}
=
{\rm Tr} \Big[ \tfrac{1}{I + A} \, {\dot A} \Big]
\; , \nonumber \\
& \!\!\!=\!\!\! &
{\rm Tr} \Big[ \big( I - A + A^2 - A^3 + A^4 - \dots \big) {\dot A} \Big]
\; , \nonumber \\
& \!\!\!=\!\!\! &
+ \, {\rm Tr} \Big[ {\dot A} + A^2 {\dot A} + A^4 {\dot A} + \dots \Big]
- {\rm Tr} \Big[ A {\dot A} + A^3 {\dot A} + A^5 {\dot A} + \dots \Big]
\; , \qquad \label{gamma dotA}
\end{eqnarray}
supplemented with the assumption of pure de Sitter 
geometry for which $A_{ij} = {\mathcal A}_{ij}$, gives:
\footnote{The VEV of the odd series vanishes because 
$\mathcal{A}_{ij}$ is a free field and its odd powers 
contain odd numbers of creation and destruction operators.}
\begin{eqnarray}
\big\langle \gamma^{ij} \dot{\mathcal{A}}_{ij} \big\rangle 
& \!\!\!=\!\!\! &
0 - \big\langle \, {\rm Tr} \big[ \mathcal{A} \dot{\mathcal{A}} 
+ \mathcal{A}^3 \dot{\mathcal{A}} 
+ \mathcal{A}^5 \dot{\mathcal{A}} + \dots \big] \big\rangle
\; , \nonumber \\
& \!\!\!=\!\!\! &
- \frac{d}{dt} \, \big\langle \, {\rm Tr} \big[ 
\, \tfrac12 \mathcal{A}^2 + \tfrac14 \mathcal{A}^4 
+ \tfrac16 \mathcal{A}^6 + \dots \big] \big\rangle
\; , \nonumber \\
& \!\!\!=\!\!\! &
- \tfrac12 \frac{d}{dt} \, \big\langle \, {\rm Tr} \big[
\ln ( 1 \!-\! A^2 ) \big] \big\rangle
\; . \label{gamma dotcalA}
\end{eqnarray}

Concerning the physical late time behaviour (\ref{Hlate}) 
of the observable $\cal{H}$, we first note from 
(\ref{N}, \ref{Cfinal3}) that $C$ is driven towards negative 
values with a lower bound of  $-1$ since at that value the 
lapse function $N$ diverges; thus $\frac{1}{N}$ is driven 
towards zero. Secondly, we note from (\ref{gamma dotcalA}) that 
$\big\langle \gamma^{ij} \dot{\mathcal{A}}_{ij} \big\rangle$
evidently increases negatively and we conclude that the initial 
$H$ is seemingly driven towards zero.

By construction, the cosmological LLOG equations contain all 
leading logarithms and should easily produce the perturbative
results order by order. Indeed, concerning the perturbative 
correspondence limit of the late time behaviour (\ref{Hlate}) 
of $\cal H$, we see using (\ref{gamma dotcalA}) that the 
lowest {\it (1-loop)} term involves $\frac{d}{dt} 
\big\langle \mathcal{A}^2 \big\rangle \!\sim\! \kappa^2 H^2$, 
the next {\it (2-loop)} term involves $\frac{d}{dt} 
\big\langle \mathcal{A}^4 \big\rangle \!\sim\! \kappa^4 H^4 \ln(a)$,
and so forth. 

This seems to agree with the physical picture of the 
real graviton pair production out of the vacuum, a 1-loop 
process which provides the source on which gravitation will 
back-react \cite{Tsamis:1996qm, Tsamis:2011ep}. The secular 
back-reaction starts at 2-loops through the negative 
gravitational interaction energy amongst the real gravitons 
produced. The result is the aforementioned behaviour of 
$\cal{H}$ with time.
\footnote{The gravitational degree of freedom 
that senses any changes in the background is $C$ and its 
equation (\ref{Cfinal3}) implies that the lowest secular 
contribution indeed starts at 2-loops: 
$\kappa^2 H^2 \big\langle \mathcal{A}^2 \big\rangle 
\!\sim\! \kappa^4 H^4 \ln(a)$.} 

Obviously we should like to extract the full all-orders 
LLOG result from our newly derived operator equations.
We expect that such a non-perturbative result will justify 
our thesis that the gravitational back-reaction to an 
accelerating cosmology can indeed slow down the expansion
rate $\cal{H}$ and provide a {\it natural} exit from 
inflation. This thesis seems to be highly inviting because 
it is based on two eminently physical facts: 
{\it (i)} the dominant force shaping cosmological evolution
is gravity; hence there is no need for fine-tuned scalar
degrees of freedom, and {\it (ii)} gravitation exhibits 
a {\it universally} attractive behaviour and therefore can 
slowly but steadily counterbalance the opposite behaviour
of the bare $\Lambda = 3H^2$.

\section{Epilogue}

In view of the long and not always smooth history of QFT
in de Sitter spacetime
\cite{Grishchuk:1977zz,
Myhrvold:1983hx,
Ford:1984hs,
Allen:1985wd, 
Antoniadis:1986sb, 
Allen:1986tt,
Floratos:1987ek, 
Higuchi:1991tm,
Dolgov:1994cq,
Polyakov:2007mm,
Giddings:2007nu,
Perez-Nadal:2007yxe,
Burgess:2010dd,
Polyakov:2012uc,
Marolf:2012kh, 
Anderson:2013ila, 
Frob:2013ht,
Anninos:2014lwa,
Frob:2014zka,
Dvali:2014gua, 
Burgess:2015ajz, 
Brandenberger:2018fdd,
Baumgart:2019clc,
Brahma:2021mng,
Colas:2022hlq,
Cable:2023gdz, 
Burgess:2024eng,
Anninos:2024fty,
Brahma:2024yor,
Sloth:2025nan},
it is perhaps reasonable to restate what seems to be 
a {\it new tool} for analyzing quantum gravitational
dynamics in a cosmological accelerating spacetime like 
de Sitter.\footnote{Some of the above references describe
re-summation techniques mostly in scalar field models
and mostly based on the original stochastic analysis 
by Starobinksy \cite{Starobinsky:1986fx}. Some other 
references naturally attempt to use the renormalization 
group towards the same goal; it seems however that 
cosmology may not be a renormalization group flow 
\cite{Burgess:2015ajz, Woodard:2008yt}.} That tool is 
the remarkably simple form (\ref{Afinal3}-\ref{Cfinal3}) 
which the gauge fixed field equations of pure gravity take 
on when we demand their restriction to incorporating the 
leading logarithms from all orders of perturbation theory.

What is missing for the time being -- afterall these 
cosmological LLOG equations for pure gravity were only 
very recently obtained -- is a technique that explicitly 
extracts the non-perturbative information out of them. 

Another important, expected, and desirable fact is that 
the LLOG equations (\ref{Afinal3}-\ref{Cfinal3}) have a 
validity range much more extended than that of perturbation 
theory. They evade the conundrum of perturbation theory, 
which breaks down when $\kappa^2 H^2 \ln(a) \!\sim\! 1$. 
The problem for perturbation theory beyond this point is 
that the {\it unknown} higher loop corrections are of the 
same order. However, equations (\ref{Afinal3}-\ref{Cfinal3}) 
tell us precisely what these higher corrections are, so these 
equations should continue to pertain to much later times. 
We have already seen that, since $C \sim O(1)$, the LLOG 
equations retain their validity until $C$ approaches $-1$. 
Equation (\ref{Cfinal3}) implies that this breakdown 
occurs for 
$\langle \gamma^{ij} A_{ij} \rangle \!\sim\! 
\frac{1}{\kappa^2 H^2}$. Hence, the breakdown points are:
\begin{equation}
\langle \gamma^{ij} A_{ij} \rangle_{pert} 
\sim \kappa^2 H^2 \ln (a)
\quad , \quad 
\langle \gamma^{ij} A_{ij} \rangle_{\scriptscriptstyle{LLOG}} 
\sim \tfrac{1}{\kappa^2 H^2}
\; . \label{breakdowns}
\end{equation}

In one sentence, we argued that the cosmological LLOG 
equations -- by containing the leading logarithms from 
all orders of perturbation theory -- can reproduce the 
laboriously obtained leading logarithms from perturbative
loops remarkably fast, and as expected can accurately
reach much further in time -- relative to perturbation 
theory -- within the inflationary era.

It seems clear that the operator equations 
(\ref{Afinal3}-\ref{Cfinal3}) have to be further studied; 
firstly to ensure the proper correspondence limit with 
explicit perturbative computations, and secondly to 
potentially obtain the complete LLOG late time solution 
for physical quantities like the expansion rate $\cal{H}$ 
and the gravitational force due to the presence of a test 
mass. 

\vspace{0.5cm}

\centerline{\bf Acknowledgements}

This work was partially supported by Taiwan NSTC grants
113-2112-M-006-013 and 114-2112-M-006-020, by NSF grant
PHY-2207514 and by the Institute for Fundamental Theory
at the University of Florida.

\newpage

\section{Appendix: The ADM decomposition}

The spatial-temporal decomposition of the conformally 
re-scaled metric is \cite{Arnowitt:1962hi}:
\begin{eqnarray}
{\widetilde g}_{\mu\nu} 
&\!\!\!\! = \!\!\!\!&
\begin{pmatrix}
-N^2 \!+\! \gamma_{kl} N^k N^l \;&\; -\gamma_{jl} N^l \\
\\
-\gamma_{ik} N^k & \gamma_{ij} \\
\end{pmatrix}
\label{3+1lower} \\
&\!\!\!\! = \!\!\!\!&
\begin{pmatrix}
\gamma_{kl} N^k N^l \;&\; -\gamma_{jl} N^l \!&\! \\
\\
-\gamma_{ik} N^k \;&\; \gamma_{ij} \\
\end{pmatrix}
-
\begin{pmatrix}
-N \\
\\
0 \\
\end{pmatrix}
_{\!\!\!\mu}
\begin{pmatrix}
-N \\
\\
0 \\
\end{pmatrix}
_{\!\!\!\nu} 
\equiv
{\overline \gamma}_{\mu\nu} \!\!- u_{\mu} u_{\nu}
\; , \qquad \label{3+1lowerB}
\end{eqnarray}
which implies the following form for its inverse: 
\begin{eqnarray}
{\widetilde g}^{\mu\nu}
&\!\!\!\! = \!\!\!\!&
\begin{pmatrix}
-\frac{1}{N^2} & -\frac{N^j}{N^2} \\
\\
-\frac{N^i}{N^2} \;&\; \gamma^{ij} \!-\! \frac{N^i N^j}{N^2} \\
\end{pmatrix}
\label{3+1upper} \\
&\!\!\!\! = \!\!\!\!&
\begin{pmatrix}
0 & 0 \\
\\
0 \;&\; \gamma^{ij} \\
\end{pmatrix}
-
\begin{pmatrix}
\frac{1}{N} \\
\\
\frac{N^i}{N} \\
\end{pmatrix}
^{\!\!\!\mu}
\begin{pmatrix}
\frac{1}{N} \\
\\
\frac{N^j}{N} \\
\end{pmatrix}
^{\!\!\!\nu} 
\equiv
{\overline \gamma}^{\mu\nu} \!\!- u^{\mu} u^{\nu}
\; . \qquad \label{3+1upperB}
\end{eqnarray}
In (\ref{3+1lower}-\ref{3+1upperB}) $N$ is the lapse 
function, $N^i$ is the shift function, and $\gamma_{ij}$ 
is the spatial metric. The ``spatial part'' is
${\overline \gamma}^{\mu\nu}$ and the ``temporal part''
is  $u_{\mu}$.

\vspace{0.5cm}


\end{document}